 \documentclass[CEJP,PDF]{cej} % use PDF command to enable PDFLaTeX driver
\usepackage{layout}
\usepackage[OT4]{fontenc}
\usepackage{amsmath}
\usepackage{textcomp}
\usepackage{hyperref}
\usepackage{amssymb}
\newcommand{\kvec}{\mathbf{k}}
\newcommand{\p}{\partial}
\newcommand{\half}{\frac{1}{2}}
\newcommand{\be}{\begin{eqnarray}}
\newcommand{\ee}{\end{eqnarray}}

\newcommand{\mn}{{\mu\nu}}

\newcommand{\rs}{{\rho\sigma}}

\newcommand{\hmn}{h_{\mu\nu}}

\newcommand{\hmnt}{\tilde{h}_{\mu\nu}}
\newcommand{\hmnupt}{\tilde{h}^{\mu\nu}}

\newcommand{\Ltri}{\mathcal{L}^{(3)}}
\newcommand{\huab}{h^{\alpha\beta}}

\newcommand{\nn}{\nonumber}
\newcommand{\humn}{h^{\mu\nu}}

\title{Massive spin-2 theories}

\articletype{}

\author{Sarah Folkerts\inst{1}\email{Sarah.Folkerts@physik.uni-muenchen.de},
        Cristiano Germani\inst{1}\email{cristiano.germani@lmu.de} and 
        Nico Wintergerst\inst{1}\email{nico.wintergerst@physik.uni-muenchen.de}}

\institute{
     \inst{1} Arnold Sommerfeld Center, Ludwig-Maximilians-Universit\"at, Theresienstr. 37, 80333 M\"unchen, Germany
    }

\abstract{
We give an introduction to massive spin-2 theories and the problem of their non-linear completion. We review the Boulware-Deser ghost problem and two ways to circumvent classic no-go theorems. In turn, massive spin-2 theories are not uniquely defined. In the case of truncated theories, we show that the Boulware-Deser ghost may only be avoided if the derivative structure of the theory is not tuned to be Einsteinian.
}

\keywords{Effective field theories, massive spin-2 theories}
\begin{document}
\maketitle

%% ###################################################################

\section{Introduction}
If low energy physics is described by the language of effective field theory (EFT), an important question to ask is what (interacting) degrees of freedom can in principle be used on four-dimensional Minkowski space.

In a Poincar\'e invariant theory, such as an EFT constructed on a four-dimensional flat space, different degrees of freedom as well as their corresponding one-particle states may be labelled by their mass and spin, which are Casimir operators of the Poincar\'e group. 

This classification is of course most useful when the one-particle states under consideration are eigenstates of the full Hamiltonian. However, in almost all cases diagonalization of the interacting Hamiltonian is extremely involved. Therefore, one introduces the concept of \emph{asymptotic states}, eigenstates of the quadratic part of the Hamiltonian.
This is due to the fact that a typical measurement involves a scattering experiment where the experimental apparatus may only monitor the {\it in} and {\it out} states far away from the scattering interactions. This brings about the concept of the S-matrix
\be
P({\text{in}, \text{out}})=\langle {\text{in}}|S|{\text{out}}\rangle\ ,
\ee
where $P\leq 1$ is the probability that the {\it in} asymptotic (far away from the scattering process) state scatters into the {\it out} asymptotic state where, importantly, {\it in} and {\it out} states are eigenstates of the {\it free} Hamiltonian. 
In almost all cases, these states are also eigenstates of the Casimir operators of the Poincar\'e group, thus allowing the above classification on the asymptotic states.
Here, we will follow this trend.

Fundamental theories must have that for all processes $P\leq 1$. This condition is what is commonly called the unitarity bound.
In an EFT instead, the considered action is only an approximation to the full quantum effective action; it corresponds to the first terms of a perturbative expansion in terms of the dimensionless parameters $E/\Lambda_s$ and $\phi/\Lambda_s$, where $E$ is the energy, $\phi$ represents any field content and $\Lambda_s$ is the so-called cut-off or strong coupling scale. Of course, if $\Lambda_s\rightarrow \infty$, the theory is valid up to any energy scale and it is called {\it renormalizable}.

Whenever the perturbative expansion breaks down, for sufficiently high transfer-energies $E>\Lambda_s$, one obtains $P > 1$. In this case, if one insists on this theory still being described in terms of a perturbative expansion, it must be rewritten in terms of new degrees of freedom materializing at an energy scale $E>\Lambda_s$. This is what is commonly called a Wilsonian UV completion.

Theories with massive and massless spin 0 and spin 1/2 fields together with massless spin 1 fields can be constructed without violation of unitary. This is, in fact, the content of the Standard Model of particle physics.

Any other degrees of freedom will inexorably violate unitarity at some energy scale (non-renormalizable theories). If it was not for gravity, a massless spin 2 particle, we could just ignore these other degrees of freedom. However, as gravitational interactions {\it do} exist, one may ask whether other non-renormalizable degrees of freedom also exist up to energy scales that are not experimentally probed yet.

A self-interacting massive vector field with mass $m$ has a typical strong coupling scale $\Lambda\sim m$. The violation of unitarity is due to the fact that the longitudinal mode of the vector, which is non-physical in the massless case, has a polarization $\vec \epsilon\propto\frac{\vec k}{m}$. Thus, the larger is the mass, the later we need to postpone the completion necessary to restore unitarity. 

This can be understood by (schematically) considering the interaction in Fourier space
\begin{equation}
\int d^4 x A_\mu J^\mu \sim \int dt\,d^3k J_{\mu;(\mathbf{k})} \sum_{\lambda = 1}^3 \epsilon^\mu_{\mathbf{k},\lambda} A_\mathbf{k}\, ,
\end{equation}
where the sum extends over the three polarizations defined by the conditions
\be
\mathbf{k}_\mu\epsilon^\mu_{\mathbf{k},\lambda} =0\ ,\cr
\epsilon_\mu{}_{;\mathbf{k},\lambda}\epsilon^\mu_{\mathbf{k},\lambda}=1\ .
\ee
The spatial part of the longitudinal polarization is defined to be parallel to the three-momentum, i.e. $\epsilon^i_{\mathbf{k},3}\propto\mathbf{k}^i$.
Explicitly, it takes on the form
\begin{equation}
\epsilon^\mu_{\mathbf{k},3} = \left(\frac{|\mathbf{k}|}{m}, \frac{\mathbf{k}}{|\mathbf{k}|}\frac{E_\mathbf{k}}{m}\right)\,,
\end{equation}
thus giving rise to a vertex 
\begin{equation}
J_{\mu;(\mathbf{k})} \epsilon^\mu_{\mathbf{k},3} A_\mathbf{k}
\end{equation}
which becomes strong at energies of order $m$.

Note that for large $|\kvec|$,
\be
\epsilon^\mu_{\mathbf{k},3} \approx \frac{1}{m} k^\mu \,.
\label{eq:longpollargek}
\ee
This signals a straightforward way to avoid the violation of unitarity. If the source $J_\mu$ is conserved, $\p_\mu J^\mu = 0$, in the large momentum limit $J_{\mu;(\mathbf{k})} \epsilon^\mu_{\mathbf{k},3} \to 0$.

We can understand this effect in a different way. We can construct one particle states by defining creation and annihilation operators $a_\kvec^\lambda$, generating the eigenstates of the non-interacting Hamiltonian. We write
\be
A_\mu^\lambda=\int \frac{d^3k}{\sqrt{2E_{\mathbf{k}}}} \epsilon^\lambda_\mu(\kvec) \left(a_\kvec^\lambda e^{i\kvec\cdot \mathbf{x}} + a_\kvec^{\dagger\lambda} e^{-i\kvec\cdot \mathbf{x}}\right)\ .
\ee
At the same time we can define a new scalar field $\phi$ as
\be
\phi=\int \frac{d^3k}{\sqrt{2E_{\mathbf{k}}}}\left[\left(i a_\kvec^{(3)}\right) e^{i\kvec\cdot \mathbf{x}} + \left(i a_\kvec^{(3)}\right)^\dagger e^{-i\kvec\cdot \mathbf{x}}\right]\ .
\ee
On an asymptotic state $|\kvec\rangle$ of four-momentum $k^\mu$ one has
\be
\partial_\mu\phi|\kvec\rangle=\frac{1}{\sqrt{2E_{\mathbf{k}}}} k_\mu \left(a_\kvec^{(3)} e^{i\kvec\cdot \mathbf{x}} + a_\kvec^{\dagger (3)} e^{-i\kvec\cdot \mathbf{x}}\right)|\kvec\rangle\ .
\ee
Henceforth
\be\label{equivpol}
\left(A_\mu^{(3)}-\frac{1}{m}\partial_\mu\phi\right)|\kvec\rangle\underset{\kvec\gg m}{\sim}\frac{m}{|\kvec|}\hat{k}_\mu a_\kvec^{(3)}|\kvec\rangle\underset{\kvec\rightarrow \infty}{\rightarrow} 0\ .
\ee
Here, $\hat{k}_\mu$ is the unit vector pointing in the direction of $\kvec$.
We thus see that the longitudinal polarization in the high energy limit is well described by a scalar field up to corrections ${\cal O}\left(\frac{m}{\kvec}\right)$. This is the essence of the St\"uckelberg decomposition of the massive vector. There, the field is decomposed into a massless vector and a scalar where the scalar is nothing else than the re-incarnation of the gauge direction of the massless case. 

To be precise, in the massless case the action is invariant under the gauge transformation
\be\label{U1}
A_\mu\rightarrow A_\mu+\frac{\partial_\mu \phi}{m}\ ,
\ee
for any $\phi$ and some mass scale $m$. In the massive case however $\phi$ represents the extra polarization at high energies as in \eqref{equivpol}. In this case, we can decompose
\be
A_\mu=\tilde A_\mu+\frac{\partial_\mu\phi}{m}\ .
\ee
As a consequence, $A_\mu$, and hence any action constructed from it, is invariant under transformations of $\tilde A_\mu$ of the type \eqref{U1} (U(1)) if the change is absorbed by a shift in the scalar $\phi$. 
We see that that the interaction of the scalar degrees of freedom to external conserved sources $J^\mu$ is absent: 
\be
\int d^4 x A_\mu J^\mu=\int d^4 x \tilde A_\mu J^\mu+\frac{1}{m}\ {\rm boundary}\ .
\ee
Thus a massive linearly interacting vector can exist without unitarity problems. The would be strong coupling scale appears in fact only in the boundary term. 

The question is now whether interacting higher spins $s\geq 3/2$ can be consistently constructed. First of all, one notes that, because of the non-trivial tensorial structure of $s\geq 3/2$ fields, all interactions must be unitarity violating \cite{Weinberg:1980kq}. Furthermore, in the case of massive fields, the extra longitudinal polarization does not decouple in the massless limit even if the field interacts with a conserved source. This is simply due to the fact that these extra polarizations always carry contributions which are not proportional to the four-momentum even in the high-energy limit.

Let us take as an example a massive spin 2 field $h_{\alpha\beta}$.
Similar to the massive vector field discussed above, the properties of this theory can also be investigated through a helicity or \emph{linear} St\"uckelberg decomposition. 

The decomposition of the field into helicity eigenmodes is in complete analogy to the massive vector. For high energies the helicity-1 component (or vectorial polarization) can be described by the derivative of a Lorentz vector ($A_\mu$), whereas the helicity-0 component (or longitudinal polarization) can be described by a scalar field $\chi$.% \cite{ANS,Kurti}. 
 The helicity-2 component (the transverse polarization) is described by a tensor $\tilde h_{\mu\nu}$. 

One then decomposes the massive spin-2 field as 
\be 
\label{eq:decomp}
\hmn = \hmnt + \frac{\p_{(\mu}A_{\nu)}}{m} + \frac{1}{3} \left(\frac{\partial_\mu \partial_\nu \chi}{m^2} + \frac{1}{2} \eta_{\mu\nu} \chi \right) \; .
\ee
Here we used the symmetrization convention $a_{(\mu}b_{\nu)}=\half(a_\mu b_\nu +a_\nu b_\mu)$.

Similar to the massive vector, interactions of the longitudinal polarization violate unitarity. 
The coupling to the energy momentum tensor is of the form (recall that the energy momentum tensor must be of dimension four)
\be
\frac{1}{\Lambda_s}\int d^4 x h_{\alpha\beta} T^{\alpha\beta}=\frac{1}{6\Lambda_s}\int d^4x \chi T+\ldots\ ,
\ee
and violates unitarity at the scale $\Lambda_s$.

If we were instead considering a non-conserved source $J^{\mu\nu}$ of mass dimension $d\geq 2$ \footnote{For dimension $d<2$ we just have kinetic or mass mixing.} we would have in addition the interaction
\be
\int d^4 x h_{\alpha\beta} J^{\alpha\beta}=\frac{\Lambda^{3-d}}{3 m^2}\int d^4x \partial_{\mu\nu}\chi \tilde J^{\mu\nu}+\ldots\ ,
\label{eq:source}
\ee 
where $\Lambda$ is the mass scale of the source $J$ and $\tilde J=J/\Lambda^{3-d}$ is the dimensionless source. This interaction creates a strong coupling scale $\Lambda_s\equiv (\frac{3 m^2}{\Lambda^{3-d}})^{\frac{1}{d-1}}$. %proportional to the mass $m$.

What we thus see is that theories describing interactions of fields of spin $s\geq 3/2$, as EFTs on a Minkowski background, cannot be fundamental (unless free) and must be UV completed. For a different route to UV completion see, e.g, \cite{classicalisation,BHselfcomp,transplanck}. 

A Poincar\'e invariant consistent theory of massless spin-2, at least at the lowest momentum expansion, must be Einstein's theory \cite{masslessspin2GR,ogievetsky, deserGR}. Therefore, at least in the low momentum limit, a massless spin-2 theory is unique up to a strong coupling scale $\Lambda_s$. The question we would like to address is whether a similar "uniqueness" theorem holds for a massive spin 2 at least up to a generic strong coupling scale $\Lambda_s$. As we shall demonstrate, this is not the case.

Before going to the next section we would like to open a parenthesis here: In case of considering a massive spin-2 theory as an infrared modification of gravity, the strong coupling can have various phenomenological consequences. 

The coupling of the longitudinal polarization $\chi$ in Eq.(\ref{eq:source}) survives in the massless limit. 
In the linear approximation, the gravitational attraction between two static bodies is stronger than in GR no matter how small the mass. While one could cure this problem by choosing $\Lambda_s \neq M_p$ in Eq.(\ref{eq:source}), this would induce changed predictions for light bending, since the energy-momentum tensor of the photon is traceless. This property is known as the vDVZ discontinuity \cite{vdvz}\footnote{Note that this problem does not persist in asymptotically de Sitter or Anti-de Sitter space. For a nonvanishing cosmological constant, the limit is indeed smooth \cite{vdvzlambda}.}. However, it was realized that the linear approximation breaks down at a distance from the source proportional to the inverse mass, the so-called Vainshtein radius. In turn, taking nonlinearities into account yields an effective source-source coupling which can be phenomenologically acceptable \cite{vainshtein}. In terms of $h_{\mu\nu}$, this can be understood from the $m^{-2}$ and $m^{-4}$ contributions to the propagator from the longitudinal mode (cf. Eq.(\ref{eq:longpollargek})) \cite{deffayet}. In terms of helicities, it can be associated to the nonlinearities of $\chi$ that become strong at the Vainshtein radius \cite{nima}. 
The precise mechanism that suppresses the fifth force on sufficiently small scales depends on the theory at hand. It may be due to the exchange of a ghost mode \cite{deffayet_vainshghost} or due to an enhancement of the kinetic term on the induced background. There, canonically normalizing its fluctuations leads to suppression of the coupling to the source \cite{dgp_vainshtein}. More on the Vainshtein mechanism, its application and viability in massive gravity theories and matching of solutions can be found, for example, in \cite{Kurti, Babichev:2013usa} (see also references therein).

\section{Massive spin-2 without self-interactions}
The action of a free massive spin-2 particle is given by what is commonly called the Fierz-Pauli action and is unique \cite{FP}.

Its construction can be understood most easily by considering the aforementioned helicity or St\"uckelberg decomposition. Demanding the absence of higher derivatives, which signal the appearance of new degrees of freedom, removes all arbitrariness in the action; only the Fierz-Pauli form allows for this property. It is given by
\label{sec:FPaction}
\begin{eqnarray}
\label{eq:FPlag}
S=\int\!d^4x\mathcal{L}&=&\int\!d^4x \left(\p_\mu h^{\mu\nu}\p_\nu h - \p_\mu h^{\rho\sigma}\p_\rho h^\mu_\sigma+\half\p_\mu h^{\rho\sigma}\p^\mu h_{\rho\sigma}-\half\p_\mu h \p^\mu h\right.\nonumber\\
&&\left. - \half m^2(h^{\mu\nu}h_{\mu\nu}-h^2)\right) ,
\end{eqnarray}
where $h\equiv h^\mu_\mu$.
 
Inserting \eqref{eq:decomp} into the quadratic action \eqref{eq:FPlag} leads to
\begin{eqnarray}\label{eq:Lheldec}
{\cal L}_\text{PF} &=& \hmnupt {\cal E}^{\rs}_{\mn} \tilde{h}_\rs - \frac{1}{8} F_\mn F^\mn + \frac{1}{12}\chi\Box\chi - \frac{1}{2} m^2 \left( \hmnupt\hmnt - \tilde{h}^2 \right)  + \frac{1}{6}m^2\chi^2 \nn \\
&& + \frac{1}{2}m^2\chi\tilde{h} + m\left(\tilde{h}\partial_\mu A^\mu - \hmnupt\partial_\mu A_\nu\right) + \frac{m}{2} \chi \partial_\mu A^\mu \; ,
\end{eqnarray}
where $\hmnupt {\cal E}^{\rs}_{\mn} \tilde{h}_\rs= \p_\mu \tilde{h}^{\mu\nu}\p_\nu \tilde{h} - \p_\mu \tilde{h}^{\rho\sigma}\p_\rho \tilde{h}^\mu_\sigma+\half\p_\mu \tilde{h}^{\rho\sigma}\p^\mu \tilde{h}_{\rho\sigma}-\half\p_\mu \tilde{h} \p^\mu \tilde{h}$ describes the linear part of the Einstein action.
For $k^2 \gg m^2$, the action becomes diagonal in field space. The individual kinetic terms for $\hmnt$ and $A_\mu$ correspond to massless linearized Einstein and Maxwell theory, respectively. Thus, in the limit where the mixing of the individual fields can be neglected, $\hmnt$ carries precisely the two helicity-2, $A_\mu$ the two helicity-1 and $\chi$ the single helicity-0  degrees of freedom. 

Note that requiring the diagonalization of the kinetic term fixes the relative factor of $1/2$ between the $\chi$-terms in \eqref{eq:decomp}. Similarly, the factors of $m$ in \eqref{eq:decomp} normalize the kinetic terms. The coefficient of the kinetic term for $\chi$ is determined by the coupling of $\hmn$ to sources: $\int d^4x T^{\mu\nu}\hmn$. The propagator of a massive spin-2 field $\hmn$ between two conserved sources $T_{\mu\nu}$ and $\tau_{\mu\nu}$ is given by 

\begin{eqnarray}
T^{\mu\nu}D_{\mu\nu,\rho\sigma}\tau^{\rho\sigma} &=& T^{\mu\nu}\frac{\left(\eta_{\mu\rho}\eta_{\nu\sigma}+\eta_{\mu\sigma}\eta_{\nu\rho}-\frac{2}{3}\eta_{\mu\nu}\eta_{\rho\sigma}\right)}{p^2-m^2}\tau^{\rho\sigma} \nonumber\\
&=&T^{\mu\nu}\frac{\left(\eta_{\mu\rho}\eta_{\nu\sigma}+\eta_{\mu\sigma}\eta_{\nu\rho}-\frac{1}{2}\eta_{\mu\nu}\eta_{\rho\sigma}\right)}{p^2-m^2}\tau^{\rho\sigma} + T^{\mu\nu}\frac{1}{6}\frac{\eta_{\mu\nu}\eta_{\rho\sigma}}{p^2-m^2}\tau^{\rho\sigma}\; .
\end{eqnarray}
The first term in the last line corresponds to the helicity-2 state $\hmnt$. The second term is an additional interaction from the extra scalar degree of freedom $\chi$ and fixes the overall normalization of it in our helicity decomposition. By considering non-conserved sources one can accordingly fix the normalization of $A_\mu$ in \eqref{eq:decomp}.

For $m=0$, the action \eqref{eq:FPlag} describes linearized Einstein gravity and is invariant under linearized diffeomorphisms, 
\be\label{gaugeinv}
h_{\mu\nu}\to h_{\mu\nu}+\half(\p_\mu \xi_\nu+\p_\nu\xi_\mu)\ , 
\ee
where $\xi_\mu(x)$ defines the linear coordinate transformation. The gauge redundancy fixes the relative coefficients of the two-derivative terms. Since both vector and scalar appear with derivatives in the St\"uckelberg decomposition, the only way for their equations of motion to be second order is for these derivative terms to drop out from the two-derivative kinetic term for $h_{\mu\nu}$. In other words, we impose on the kinetic part of the Lagrangian the condition 
\be\label{gauge}
{\cal L}(h_{\mu\nu})={\cal L}(h_{\mu\nu}+\partial_{(\mu}\tilde A_{\nu)}+\partial_{\mu\nu}\tilde\chi)+{\rm boundaries}\ ,
\ee
where $\tilde A_\mu$ and $\tilde \chi$ are respectively a vector and a scalar.
This is equivalent to the gauge invariance \eqref{gaugeinv} for a specific $\xi_\mu$.

The uniqueness of said structure can also be understood from a Hamiltonian analysis. Let us first examine the kinetic term.
After having integrated by parts such that $h_{00}$ and $h_{0 i}$ do not appear with time derivatives, the canonical momenta of the Lagrangian \eqref{eq:FPlag} are 
\be\label{eq:canmomFP}
\pi_{i j}=\frac{\p\mathcal{L}}{\p \dot{h}_{i j}}=\dot{h}_{i j}-\dot{h}_{i i} \delta_{i j}-2\p_{(i}h_{j)0} .
\ee
 The other canonical momenta ($\pi_{0 0}$ and $\pi_{0 i}$) are zero due to the integration by parts. Inverting \eqref{eq:canmomFP}, one obtains 
 \be\label{eq:velFP}
 \dot{h}_{i j}=\pi_{i j}-\pi_{k k}\delta_{i j}+2\p_{(i}h_{j)0} .
 \ee
Performing the Legendre transformation and rewriting the Lagrangian in terms of the canonical momenta yields
\begin{eqnarray}\label{eq:FPacwicanmom}
\mathcal{L}&=&\pi_{i j}\dot{h}_{i j} - \mathcal{H} +2h_{0 i}\p_j\pi_{i j} +h_{0 0}(\nabla^2h_{i i}-\p_i\p_j h_{i j})\; , \nonumber\\
\mathrm{where}\quad\mathcal{H}&=&\half \pi_{i j}^2-\frac{1}{4}\pi_{i i}^2+\half\p_k h_{i j}\p_k h_{i j}-\p_i h_{j k}\p_j h_{i k}+\p_i h_{i j}\p_j h_{k k} - \half \p_i h_{j j} \p_i h_{k k} .\nn\\
\end{eqnarray}
The canonical momenta for $h_{00}$ and $h_{0 i}$ are zero and the variables themselves appear only linearly in terms without time-derivatives. They are Lagrange multipliers which give the constraint equations $\nabla^2h_{i i}-\p_i\p_j h_{i j}=0$ and $\p_j \pi_{i j}=0$. All these constraints commute, in the sense of Poisson brackets, with each other. Hence, the constraints are first class (for an introduction to constrained systems see for example \cite{Dirac, Henneaux}). This is characteristic for theories with a gauge symmetry. The constraints together with the gauge transformations reduce the physical phase space to a four dimensional hypersurface, which is described by the canonical coordinates of the two physical polarizations of the massless spin-2 graviton and their conjugate momenta.

Adding a mass term to the analysis changes the Hamiltonian and the Lagrangian of \eqref{eq:FPacwicanmom} in the following way

\begin{eqnarray}\label{eq:FPacwicanmom}
\mathcal{L}&=&\pi_{i j}\dot{h}_{i j} - \mathcal{H} +m^2 h^2_{0 i}+2h_{0 i}\p_j\pi_{i j} +h_{0 0}(\nabla^2h_{i i}-\p_i\p_j h_{i j}-m^2 h_{i i}) \; ,\nonumber\\
\mathrm{where}\quad  \mathcal{H}&=&\half \pi_{i j}^2-\frac{1}{4}\pi_{i i}^2+\half\p_k h_{i j}\p_k h_{i j}-\p_i h_{j k}\p_j h_{i k}+\p_i h_{i j}\p_j h_{k k}\nonumber\\
&& - \half \p_i h_{j j} \p_i h_{k k} +\half (h_{i j}h_{i j}-h_{i i}^2).
\end{eqnarray}
Note that the conjugate momenta are unchanged by the additional mass term. However, the structure of the Lagrangian is different and $h_{0i}$ is no longer a Lagrange multiplier. Nevertheless, it is still non-dynamical and its equation of motion yields the algebraic relation

\be\label{eq:h0ieom}
h_{0 i}=-\frac{1}{m^2}\p_i \pi_{i j} .
\ee
$h_{00}$ still is a Lagrange multiplier and it enforces the constraint 
\be\label{eq:consh00}
\nabla^2h_{i i}-\p_i\p_j h_{i j}-m^2h_{i i}=0
\ee
 which is now of second class. 
Requiring that the constraint is conserved in time, i.e. that it commutes with the Hamiltonian, gives rise to a secondary constraint. Since $h_{0i}$ is determined by \eqref{eq:h0ieom} and $h_{00}$ gives two second class constraints (one primary and one secondary), the resulting physical phase space is then ten dimensional describing the five physical polarizations of the massive spin-2 particle and their conjugate momenta. Departing from the Fierz-Pauli mass term introduces nonlinearities in $h_{00}$ and the constraint which fixes the trace $h_{i i}$ to zero is lost resulting in either a tachyonic or ghost-like sixth degree of freedom \cite{BD,vanNieuwenhuizen}.
 
Let us briefly mention coupling to sources. Adding a source term to the Lagrangian \eqref{eq:FPlag} of the form $h_{\mu\nu}T^{\mu\nu}$ does not change the linear constraint analysis. No matter whether the source is conserved, $\p_\mu T^{\mu\nu}=0$, or not, the source coupling will only introduce $h_{00}$ and $h_{0 i}$ linearly and without time derivatives and therefore it will not affect the number of constraints. Note that this holds true for any linear coupling of $h_{\mu\nu}$ to sources.

\section{Self-interacting theories}
 
We now focus on the question of self-interactions in massive spin-2 theories. We address subtleties in the construction and inquire whether uniqueness theorems can exist similar to the
case of the Einstein theory for a massless spin-2 field.

\subsection{Boulware-Deser ghost}

Boulware and Deser (BD) argued in \cite{BD} that simply introducing a mass term for the full nonlinear theory of general relativity 
reintroduces the sixth degree of freedom which could be tuned away in the Fierz-Pauli theory. Although this result turned out to be not generic, it is instructive to see their reasoning.

Let us first consider pure general relativity. Using the ADM formalism \cite{ADM} in which a general metric can be re-written as
\be\label{eq:ADM}
ds^2=g_{\alpha\beta}dx^\alpha dx^\beta=-N^2dt^2+\gamma_{ij}\left(dx^i+N^idt\right)\left(dx^j+N^jdt\right)\ ,
\ee
where $\gamma_{i j}\equiv g_{i j}$, $N\equiv (-g^{00})^{-\half}$ (lapse), $N_i\equiv g_{0 i}$ (shift). The full action reads (for simplicity we set the Planck mass to one)
\be\label{eq:GRLag}
S=\int\!d^4x \sqrt{-g} \;R=\int\!d^4x(\pi_{i j}\dot{\gamma}_{i j}-N R^{(0)}-N_i R^i-2(\pi^{i j}N_j-\half \pi N^i+N^{|i}\sqrt{\gamma})_{|j}) ,
\ee
All curvatures are functions of $\gamma_{i j}$ and $\pi_{i j}$, but do not depend on $N$ or $N_i$. R is the four dimensional Ricci scalar and $-R^{(0)}\equiv \;^{3}\! R +\gamma^{-\half}(\half\pi^2-\pi_{i j}\pi^{i j})$ and $^{3}\! R $ is the three dimensional Ricci scalar with respect to the metric $\gamma_{i j}$.  $R^i=-2\pi^{i j}_{ | j}$, where the bar ``$_|$'' denotes covariant differentiation with respect to the spatial metric $\gamma_{i j}$. 

In the massless theory, $N$ and $N_i$ are Lagrange multiplier which enforce first class constraints on the system, thereby eliminating four (and correspondingly eight phase space) degrees of freedom yielding 2 propagating helicities of the massless spin-2 particle. We now introduce the Minkowski background by expanding
\be
g_{\alpha\beta}=\eta_{\alpha\beta}+h_{\alpha\beta}\ ,
\ee
where $\eta_{\alpha\beta}$ is the Minkowski metric and $h_{\alpha\beta}$ is a tensor on the flat background. Its indices are consequently raised and lowered by the Minkowski metric. The inverse metric $g^{\alpha\beta}$ is given by an infinite series of $h_{\alpha\beta}$ and can be obtained from $g^{\alpha\mu}g_{\mu\beta} = \delta^\alpha_\mu$.
At linear order $N=1-\half h_{0 0}$ and $N_i=h_{0 i}$ and one recovers the result of the previous section. At nonlinear order, however, 
\be
N^2=(1-h_{00})-h_{0i} h_{0j} g^{ij}\ ,
\ee
whereas $N_i$ remains unchanged.

The Fierz-Pauli mass term $f=(h_{\mu\nu}h^{\mu\nu}-h^2)$ can nevertheless easily be expressed in terms of $N_i$ and the nonlinear $N$ \cite{BD},

\be\label{eq:FPADM}
f=h_{i j}^2-h_{i i}^2-2N_i^2+2h_{i i}(1-N^2-N_i N^i) .
\ee
In contrast to the linear case, here $N$ (which to linear order is equivalent to $h_{0 0}$)  appears quadratically in the mass term although still appearing linearly in the full non-linear derivative (Einstein) structure of the theory. 
Therefore now neither $N$ nor $N_i$ are Lagrange multipliers.

Thus, at the full non-linear level, the trace $h_{i i}$ is no longer constrained since the constraint was related to the fact that $N$ was a Lagrange multiplier. 
Therefore, there are six degrees of freedom propagating: The so-called Bouleware-Deser ghost propagates on top of the five degrees of freedom of the Fierz-Pauli massive spin-2. We will see that this conclusion, although correct generically, can be avoided for specific theories.

The simplest example is the free Fierz-Pauli theory discussed above. There, since the expansion is truncated at the linear level we have that $N=1-\half h_{00}$ and the mass term in the Lagrangian is
\be
f=h_{i j}^2-h_{i i}^2-2N_i^2+2h_{ii}(1-2N).
\ee
Thus, as in the derivative part of the action (the linearized Einstein-Hilbert Lagrangian), $N$ only appears linearly. In other words, the lapse is here again a Lagrange multiplier, as in General Relativity. 

The philosophy of avoiding the BD ghost will be the same for interacting theories: we will search for theories that can be written in terms of a linear lapse function acting as a Lagrange multiplier. In order to do that and to avoid the BD conclusions, we will have to either deform the derivative structure of the massless theory and/or the non-derivative structure.

\subsection{Cubic interactions for a massive spin-2 particle}

We will start by considering the simplest possible interaction, a cubic interaction as described in \cite{ANS}.

There, the idea was to consider a cubic interaction that keeps the structure of the \emph{linear} Fierz-Pauli action. In other words, by deviating from the Einsteinian derivative structure at the cubic order, $ N =1-\half h_{00}$ remains a Lagrange multiplier also in the nonlinear theory. Non-derivative interactions can then be constructed that preserve this property.

This construction can straightforwardly be achieved by considering the most general cubic interaction with at most two derivatives on $h_{\mu\nu}$. Demanding linearity in $h_{00}$ fixes all coefficients besides respective prefactors of the zero- and two-derivative terms.

The unique structure is found to be \cite{ANS}
\begin{eqnarray}
\label{eq:lastcoe}
\Ltri&=& \frac{k_1}{\Lambda^7}\big(\huab\p_\alpha\humn\p_\beta\hmn- \huab\p_\alpha h\p_\beta h +4\huab\p_\beta h \p_\mu h_\alpha^\mu-2 \humn \p_\alpha h\p^\alpha\hmn+h \p_\mu h \p^\mu h\nonumber\\
&&-3\humn\p_\alpha h^\alpha_\mu \p_\beta h_\nu^\beta-4\humn\p_\nu h_\mu^\alpha \p_\beta h_\alpha^\beta +3 h \p_\mu \humn\p_\alpha h_\nu^\alpha+ 2\humn \p^\alpha\hmn\p_\beta h_\alpha^\beta\nonumber\\
&&-2h\p_\alpha h \p_\beta \huab+ \humn \p_\alpha h_{\nu\beta}\p^\beta h_\mu^\alpha+2\humn \p_\beta h_{\nu\alpha}\p^\beta h_\mu^\alpha-h\p_\alpha \hmn \p^\nu h^{\mu\alpha}-h\p_\alpha \humn \p^\alpha\hmn\big) \nn \\
&&+ \frac{k_{2}}{\Lambda^5}\big(2 h_\nu^\mu h^\nu_\rho h^\rho_\mu-3  h h_{\mn}\humn+h^3\big) \; .
\end{eqnarray}

In terms of the components of $h_{\mu\nu}$, for example, the non-derivative part is given by 
\be\label{eq:3rdordcompnD}
\mathcal{L}=\frac{3 k_{2}}{2}h_{00}(h_{i i}^2-h_{i j}^2)+\mbox{terms independent of $h_{00}$} .
\ee
Hence, $h_{00}$ and $h_{0i}$ appear in the same way as in the free action.
We do not display the explicit expression for the derivative part because the expression is rather lengthy. Still, one can easily check that also there $h_{0i}$ remains non-dynamical and can be solved for algebraically, yielding 3 constraints on $\hmn$. Furthermore, $h_{00}$ appears as a Lagrange multiplier in \eqref{eq:lastcoe} and accordingly eliminates another two degrees of freedom \cite{ANS}.

The fact that the action \eqref{eq:lastcoe} propagates five degrees of freedom can also be checked in the helicity decomposition \eqref{eq:decomp}. 
Inserting the decomposition into \eqref{eq:lastcoe} reveals that the corresponding equations of motion are free of higher time derivatives on the helicity components.

Indeed, this is in direct correspondence to the Hamiltonian analysis outlined above. 
The components $h_{00}$ and $h_{0 i}$ are exactly those components of $\hmn$ which can introduce higher time derivatives on the equations of motion as, in terms of helicities, these correspond to $\p_0^2\chi$, $\p_0 A_0$, $\p_0\p_i\chi$ and $\p_0A_i$. Therefore, any action free of higher derivatives on the helicities, requires $h_{00}$ to be a Lagrange multiplier and $h_{0i}$ to be nondynamical.

Up to boundary terms, one can rewrite the above Lagrangian in a compact form \cite{Kurtsim} as follows

\be\label{eq:antisymL3}
\Ltri=k_1\epsilon^{\alpha_1\ldots\alpha_4}\epsilon^{\beta_1\ldots\beta_4} \p_{\alpha_1}\p_{\beta_1}h_{\alpha_2\beta_2}\ldots h_{\alpha_4\beta_4} + k_{2} \epsilon^{\alpha_1\ldots\alpha_3\sigma_4}\epsilon^{\beta_1\ldots\beta_3}_{\sigma_4}h_{\alpha1\beta_1}\ldots h_{\alpha_3\beta_3} \; .
\ee
$\epsilon^{\alpha_1\ldots\alpha_4}$ denotes the totally antisymmetric four-tensor in four dimension. From its antisymmetry properties it is then simple to conclude that the constraint structure of the free Lagrangian is preserved. If there is one $h_{00}$ in \eqref{eq:antisymL3}, then there cannot be any other factor of it in that term. Therefore, $h_{00}$ can only appear as a Lagrange multiplier. Terms with $h_{0 i}$ can carry at most one time derivative and one power of $h_{0 i}$ or only spatial derivatives and at most two powers of $h_{0 i}$; all other terms have spatial indices. Variation with respect to $h_{0 i}$, thus, leads to a constraint equation for itself which defines it algebraically in terms of the components $h_{i j}$.

\subsection{Resummed theories}
The second possible route to find nonlinear extensions of the Fierz-Pauli theory is to retain the Einsteinian derivative structure and construct a nonlinear extension of the mass term. In this case one searches for a theory preserving a similar constraint structure of the lapse for the full Einstein theory. This approach was taken in \cite{dRGT}. 

What we learned from the analysis of BD is that the lapse $N$ cannot be a lagrange multiplier if the the two following assumptions co-exist
\begin{itemize}
\item The derivative structure is Einsteinian
\item $N$ and $N^i$ are independent variables.
\end{itemize}
As we are interested in the class of theories that fulfill the former assumption, one needs to relax the latter, thereby keeping $N$ as a Lagrangian multiplier to eliminate the BD ghost. The theory with this property has been constructed in \cite{dRGT}, the so-called dRGT massive gravity.

In other words, the theory of \cite{dRGT} is a deformation of General Relativity with non-derivative term such that \cite{HassanRosen}
\begin{itemize}
\item The derivative structure is Einsteinian.
\item $N^i$ can be fully traded by a new variable $n^i(N, N^i, h_{ij})$.
\item After the field redefinition, $N$ only appears linearly in the action and without derivatives.
\end{itemize}
The second condition forces the redefinition to be of type
\be\label{eq:lapsered}
N^i=\left(\delta^i_j+N D^i_j\right)n^j\ ,
\ee
where $D^i_j$ is an appropriate matrix independent upon $N$.\footnote{Note that we would get exactly the same results without the field redefinition. This field redefinition only makes manifest that the lapse is a Lagrange multiplier.} Of course, any truncation in powers of $h_{ij}$ of this construction would bring back the BD ghost. 

For our purposes, we adopt the notation of \cite{HassanRosen}. We write the resummed theory of \cite{dRGT} in terms of the inverse metric $g^{-1}$ and an auxiliary background metric $\eta$. The action of dRGT massive gravity can then be written according to \cite{HassanRosen} (here we re-introduce the Planck mass $M_P$)
\be
\label{eq:dRGTRosen}
S=M_P^2\int d^4x\sqrt{-g}\left[R(g)+2m^2\sum_{n=0}^2\beta_n e_n\left(\sqrt{g^{-1}\eta}\right)\right]\;,
\ee
where $m$ is the graviton mass and the $e_n(\mathbb{X})$ are functions of matrix traces given by 
\begin{align}
e_0(\mathbb{X})=1\;, e_1(\mathbb{X})=[\mathbb{X}],e_2(\mathbb{X})=\half([\mathbb{X}]^2-[\mathbb{X}^2])\;.
\end{align}
The square brackets denote the trace and $\beta_0=6$, $\beta_1=-3$ and $\beta_2=1$ for dRGT massive gravity \cite{dRGT,HassanRosen}. Note that the coefficients are chosen such that the action describes a flat background without a cosmological constant. The matrix $\sqrt{g^{-1}\eta}$ is defined by $\sqrt{g^{-1}\eta}\sqrt{g^{-1}\eta}=g^{\mu\nu}\eta_{\nu\rho}$. Since $\eta_{\mu\nu}$ transforms as a rank-two tensor, the action \eqref{eq:dRGTRosen} is invariant under general coordinate transformations.  

 Expanding the action \eqref{eq:dRGTRosen} to second order in the metric perturbations $g_{\mu\nu}=\eta_{\mu\nu}+h_{\mu\nu}$, one recovers the Fierz-Pauli action \eqref{eq:FPlag}.
 
As suggested in \cite{dRGT} and later shown in \cite{HassanRosen,Hassan2,dRGTstuck,dRGThel}, the action \eqref{eq:dRGTRosen} indeed only propagates five degrees of freedom. In order to see this, one can redefine the shift as \eqref{eq:lapsered}. This has been also done for the full nonlinear action in \cite{HassanRosen}. We will briefly discuss their findings. 

A constraint analysis is most conveniently carried out by using the ADM decomposition \cite{ADM}. Using \eqref{eq:ADM}, the Lagrangian \eqref{eq:dRGTRosen} is given by
\be\label{eq:Laglapse}
M_P^{-2}\mathcal{L}=\pi^{i j}\p_t\gamma_{i j}+N R^0+ R_i N^i + 2m^2 \sqrt{\det{\gamma}}N\sum_{n=0}^2\beta_n e_n\left(\sqrt{g^{-1}\eta}\right)\;.
\ee
The mass term includes $N$ in a non-linear way and is, therefore, responsible for the seeming loss of the constraint. However, one can redefine the shift $N_i$ by \eqref{eq:lapsered} and finds the following Lagrangian \cite{HassanRosen,Hassan2}
\be\label{eq:Lagafterred}
M_P^{-2}\mathcal{L}=\pi^{i j}\p_t\gamma_{i j}-\mathcal{H}_0(\pi^{i j},\gamma_{i j},n_j)+N\mathcal{C}(\pi^{i j},\gamma_{i j},n_j)\;,
\ee
where $\mathcal{H}_0$ is the Hamiltonian and $\mathcal{C}$ is the additional constraint ensuring that only five of the six components of $\gamma_{i j}$ are propagating. 
Thus, we have established that there are three independent variables $n_i$ which are not propagating and algebraically determined by their equations of motion and there is one Lagrange multiplier $N$ which yields a constraint equation for $\pi^{i j}$ and $\gamma_{i j}$. Therefore, there are only five propagating independent degrees of freedom which constitute the massive graviton. 

It is, however, important to note that the redefinition \eqref{eq:lapsered} can only be used when considering the full non-linear action \eqref{eq:dRGTRosen}. Whenever truncating the theory, this ceases to be valid and thus one is left with six propagating degrees of freedom. 

One might be puzzled by analyzing the action \eqref{eq:dRGTRosen} in terms of the helicity decomposition \eqref{eq:decomp}. There, indeed, are higher derivatives (apparently signaling new degrees of freedom) appearing on the equations of motion of, e.g., the scalar helicity $\chi$ for the nonlinear terms \cite{ANS}. The lowest suppression scale of these terms is $\Lambda_5=(m^4M_P)^\frac{1}{5}$. In the full theory, however, this scale is redundant and can be removed by a field redefinition \cite{dRGThel}\footnote{Note that this field redefinition is in fact necessary in order to define a Hamiltonian in terms of the helicities. Without it, the relation between the canonical momenta and time derivatives of fields is not invertible. This reflects the redundancy of the coupling in the full theory.}. With this field redefinition also the higher derivative terms disappear. It can be shown that this happens for all scales below $\Lambda_3=(m^2 M_P)^\frac{1}{3}$ \cite{dRGThel}, such that the theory in terms of the redefined fields is free of higher derivative interactions. Note, however, that when truncating the theory to any finite order this is no longer the case. 

In works subsequent to \cite{dRGT,HassanRosen}, it was furthermore shown that the absence of the sixth ghost like degree of freedom can also be confirmed in the St\"uckelberg language (see for example \cite{dRGTstuck})\footnote{Note that there is still ongoing discussion in the literature about possible consistency issues of massive gravity, such as superluminality and acausality, e.g. in \cite{Dubovsky:2005xd, Gruzinov:2011sq, Burrage:2011cr, deFromont:2013iwa, deser_waldron}, and instabilities, for example in \cite{Tasinato:2012ze, DeFelice:2012mx, Kuhnel:2012gh, Babichev:2013una}.}.

\section{Conclusions}
Within this work we have addressed the question whether theories of a single interacting massive-spin 2 field obey similar uniqueness theorems as in the massless case.

For a long time, it was doubtful whether there even exists one consistent theory that describes self-interactions of a massive spin-2 particle. The fact that adding the Fierz-Pauli mass term to the Einstein-Hilbert action introduces nonlinearities in the lapse into the action was taken as the basis of a no-go-theorem for nonlinear extensions of Fierz-Pauli theory. It was argued that any such extension necessarily leads to the appearance of a sixth unphysical and ghost-like polarization in the theory, the Boulware-Deser ghost.

We have reviewed two possible ways to circumvent this apparent theorem. One is to sacrifice the Einsteinian derivative structure, such that the $(00)$-component of the tensor field $h_{\mu\nu}$ enters the action only linearly even when self-interactions are added. This ensures that only five degrees of freedom are propagating. We have shown that this property can equivalently be checked in a helicity decomposition of the massive tensor. The found action is characterized by the absence of higher derivatives on the helicity components. It is the unique theory with this property.

The second route is to leave the derivative structure untouched, but instead adding a potential for the massive spin-2 field in such a way that guarantees the presence of a Lagrange multiplier in the system. By casting the action into an appropriate form, this Lagrange multiplier is once again given by the lapse.

The latter approach, since it relies on redundancies of the full action, requires a full resummation of the theory. Any truncation to finite order appears to propagate more than five degrees of freedom. However, the scale at which this additional degree of freedom appears coincides with the scale at which nonlinearities become important. Henceforth, conclusions can only be drawn from the resummed theory.

We have further addressed the issue of higher derivatives in the helicity decomposition in the latter class of theories. While these are present, the fact that redundancies are present prevents one from constructing a Hamiltonian. A field redefinition is necessary in order to be able to invert the canonical momenta; after this redefinition, no more  higher derivatives are present. The Hamiltonian of the theory does not suffer from an Ostrogradski linear instability.

The experimental viability of either theory is unknown. The former deviates from the well probed Einsteinian cubic vertex and is therefore not viable as a massive graviton. The latter has the correct vertex structure. However, choosing its mass to be of the order of the Hubble scale leads to a strong coupling already at very low energies, $\Lambda_s=(m^2 M_P)^{1/3}\sim ({1000\ {\rm km}})^{-1}$. 

Either theory could in principle describe self-interactions of a massive spin-2 meson and could therefore be of different phenomenological interest.

\end{document}